# Experimentos sobre la Hibridación en Plantas
# Gregor Mendel

## 1. Introducción

La fecundación artificial de plantas ornamentales para producir nuevas variantes de color nos indujo a los experimentos que se analizan aquí. La notable regularidad con la que reaparecían las mismas formas híbridas cada vez que se producía la fecundación entre las mismas especies fue el estímulo para posteriores experimentos, cuyo objetivo era seguir el desarrollo de los híbridos en su progenie.

Observadores atentos como Kölreuter, Gärtner, Herbert, Lecocq, Wichura y otros han perseguido incansablemente este objetivo durante parte de sus vidas. En concreto, Gärtner, en su obra *Die Bastarderzeugung im Pflanzenreiche* [*La Producción de Híbridos en el Reino Vegetal*], ha recopilado observaciones muy valiosas; y, más recientemente, Wichura ha publicado resultados fundamentales sobre su investigación en los híbridos del sauce. Que todavía no se haya formulado con éxito una ley general que rija la formación y el desarrollo de los híbridos no sorprende a los conocedores de la extensión del tema y de las dificultades que entrañan los experimentos de esta índole. Únicamente cuando se dispongan de resultados procedentes de experimentos detallados en plantas de diversos órdenes se podrá alcanzar una conclusión firme.

Los que examinen el trabajo en este campo llegarán a la convicción de que, entre los numerosos experimentos realizados, ninguno se ha llevado a cabo en extensión y en forma tales que hagan posible determinar el número de formas diferentes que aparecen en la progenie de los híbridos, con el fin de clasificar certeramente estas formas en generaciones separadas y establecer con precisión sus relaciones estadísticas.

Ciertamente, se requiere ánimo para emprender un trabajo tan extenso; sin embargo, este parece ser el único medio adecuado para llegar finalmente a la resolución de una cuestión sobre la historia evolutiva de las formas orgánicas, cuya relevancia no debe subestimarse.

Este trabajo presenta los resultados de lo que pretende ser un experimento detallado. Como lo requería la tarea, el experimento se limitó a un grupo relativamente pequeño de plantas, y se ha concluido en lo esencial solo después del transcurso de ocho años. Si el plan según el cual se llevaron a cabo los diversos experimentos es el más apropiado para alcanzar el objetivo perseguido queda a la benévola decisión del lector.

**2. Selección de las Plantas Experimentales**

El valor y la utilidad de cualquier experimento dependen de la adecuación de los materiales al objetivo que se aborda. Por tanto, en el caso que nos ocupa, son relevantes tanto la selección de las especies de plantas para los experimentos como la manera en que estos se llevan a cabo.

La selección del grupo de plantas para experimentos de esta índole debe hacerse con sumo cuidado para evitar, desde el principio, que los resultados se cuestionen.

Las plantas experimentales de la muestra deben cumplir necesariamente los siguientes requisitos:
1. Deben poseer caracteres diferenciales constantes.
2. En el momento de la floración, sus híbridos deben estar protegidos, o debe ser fácil protegerlos, de la acción del polen de otros individuos.
3. Los híbridos y su progenie, en las generaciones sucesivas, no deben sufrir ninguna alteración apreciable en su fertilidad.

La adulteración a través del polen de otro individuo, si ocurriera inadvertidamente en el transcurso del experimento, conduciría a conclusiones completamente falsas. La fertilidad reducida o la esterilidad completa de ciertas formas, como las que aparecen en la progenie de muchos híbridos, dificultarían en gran medida los experimentos o los frustrarían por completo. Para descubrir las proporciones de las formas híbridas entre sí y con respecto a sus progenitores, parece necesario que todos y cada uno de los miembros que se desarrollan en la serie, en cada generación, se someta a observación.

Desde un principio se prestó particular atención a las *Leguminosae* por su peculiar estructura floral. Los experimentos realizados con varios miembros de esta familia llevaron a la conclusión de que el género *Pisum* cumple suficientemente los requisitos necesarios. Algunas formas completamente independientes de este género poseen caracteres constantes y fáciles de reconocer con seguridad, y dan lugar a una progenie híbrida perfectamente fértil cuando se cruzan entre sí La perturbación por el polen de otros individuos no ocurre fácilmente, ya que los órganos de fecundación están estrechamente envueltos dentro de la quilla y la apertura de las anteras se produce dentro del capullo, de modo que el estigma queda recubierto de polen antes de que se abra la flor. Así pues, esta circunstancia adquiere una acusada relevancia. Otras ventajas que merecen mencionarse son la facilidad de cultivo de estas plantas en campo abierto y en macetas, así como su período de crecimiento relativamente corto. La fecundación artificial es, sin duda, un proceso laborioso, pero casi siempre tiene éxito. Para ello, se abre el capullo antes de que esté perfectamente desarrollado, se separa la quilla y se extraen cuidadosamente todos los estambres con unas pinzas; después, el estigma puede espolvorearse con polen de otro individuo.

Se obtuvo un total de 34 variedades de guisantes, más o menos diferentes, de varios proveedores de semillas y se sometieron a una prueba de dos años. En una variedad se observaron unas pocas formas muy distintas entre un gran número de plantas idénticas. Sin embargo, al año siguiente no hubo variación entre aquellas pocas formas diferentes, que coincidieron en todos los sentidos con otra variedad obtenida del mismo proveedor de semillas; sin duda, las semillas se habían mezclado accidentalmente. Todas las demás variedades produjeron una progenie absolutamente idéntica y constante; al menos, en los dos años de prueba, no se observó ninguna variación esencial. De estas variedades se seleccionaron 22 para la fecundación cruzada, y se cultivaron anualmente durante la duración de los experimentos. Todas permanecieron constantes sin excepción.

Su clasificación sistemática es difícil e incierta. Si se aplicara la definición más estricta del término especie, según la cual solo pertenecen a la misma especie aquellos individuos que exhiben los mismos caracteres en exactamente las mismas condiciones, entonces no habría dos de esas variedades que pudieran considerarse como pertenecientes a la misma especie. No obstante, según la opinión de expertos en la materia, la mayoría pertenece a la especie *Pisum sativum*, mientras que las demás fueron consideradas y descritas, o bien como subespecies de *P. sativum*, o bien como especies independientes, como *P. quadratum*, *P. saccharatum* y *P. umbellatum*. En cualquier caso, la posición que se les pueda asignar en un sistema de clasificación carece por completo de importancia para los experimentos descritos aquí. Resulta imposible trazar una línea definida que diferencie tanto entre especies y variedades como entre los híbridos de especies y variedades.

**3. División y Distribución de los Experimentos**

Si dos plantas que difieren constantemente en uno o varios caracteres se unen por fecundación, los caracteres comunes se transmiten sin cambio a los híbridos y su progenie, como han demostrado numerosos experimentos. Sin embargo, cada par de caracteres diferentes se une en el híbrido para formar un nuevo carácter, que, generalmente, está sujeto a variaciones en la progenie del híbrido. El objetivo general del experimento es observar estas variaciones para cada par de caracteres diferentes, con el fin de formular una ley que determine cuándo ocurren dichas variaciones en las generaciones sucesivas. Este experimento, consecuentemente, se divide en tantos experimentos separados como caracteres diferenciadores constantes hay en las plantas experimentales.

Las diferentes formas de guisantes seleccionadas para la fecundación muestran diferencias en: (1) la longitud y el color del tallo; (2) el tamaño y la forma de las hojas; (3) la posición, el color y el tamaño de las flores; (4) la longitud de los pedúnculos florales; (5) el color, la forma y el tamaño de las vainas; (6) la forma y el tamaño de las semillas; y (7) el color de la cubierta de la semilla y del albumen. Ahora bien, algunos de estos caracteres no permiten una separación precisa y segura porque la diferencia reside en un "más o menos" difícil de determinar. Tales caracteres no se utilizaron para los experimentos separados, que se limitaron a los caracteres que se presentaban clara y definidamente en las plantas. El resultado de los experimentos debe mostrar si todos estos caracteres siguen un comportamiento regular en sus uniones híbridas, y si puede extraerse de estos hechos alguna conclusión respecto de los caracteres que poseen una importancia inferior.

Los caracteres que se seleccionaron para los experimentos se relacionan con:
1. Las diferencias en la forma de las semillas maduras. Estas son o bien redondas o algo redondeadas, y las depresiones, si las hay, ocurren en la superficie y son poco profundas; o bien son irregularmente angulosas y profundamente rugosas (*P. quadratum*).
2. Las diferencias en el color del albumen de la semilla (endospermo). El albumen de las semillas maduras es de color amarillo pálido, amarillo brillante o naranja; o bien posee un color verde más o menos intenso. Esta diferencia de color es evidente en las semillas, ya que sus cubiertas son transparentes.
3. Las diferencias en el color de la cubierta de la semilla. Esta es o bien de color blanco –un carácter consistentemente asociado con el color de la flor blanca–, o es gris, marrón grisáceo, o marrón cuero con o sin manchas violetas, en cuyo caso el color del estandarte parece violeta, el de las alas púrpura, y el tallo en la base de las axilas de las hojas está teñido de rojizo. Las cubiertas grises de las semillas se vuelven de color marrón negruzco en agua hirviendo.

4. Las diferencias en la forma de la vaina madura. Esta está o bien hinchada, sin estrechamientos en ciertos puntos; o bien profundamente constreñidas entre las semillas y más o menos rugosa (*P. saccharatum*).
5. Las diferencias en el color de la vaina no maduras. O bien es de color verde claro u oscuro, o bien de color amarillo brillante, un color que comparten los tallos, las nervaduras de las hojas y los sépalos.[1]
6. Las diferencias en la posición de las flores. O bien son axiales, es decir, distribuidas a lo largo del tallo; o bien terminales, esto es, reunidas en el extremo del tallo en una falsa umbela corta, en cuyo caso la parte superior del tallo tiene una sección transversal más o menos ensanchada (*P. umbellatum*).
7. Las diferencias en la longitud del tallo. La longitud del tallo es muy variable en algunas formas; no obstante, es un carácter constante en cada una de ellas, ya que las plantas sanas, desarrolladas en el mismo suelo, solo están sujetas a variaciones sin importancia de este carácter. Para poder discriminar con certeza en los experimento con este carácter, se unió un tallo más largo, de 6 a 7 pies [de 1,8 a 2,1 metros], con uno más corto, de 0,75 a 1,5 pies [de 23 a 45,7 centímetros].

Cada par de los caracteres diferenciales antes mencionados se unió a través de la fecundación. En concreto, para cada experimento se realizó el siguiente número de fecundaciones en las plantas:

- Experimento 1:     60 fecundaciones en 15 plantas.
- Experimento 2:     58 fecundaciones en 10 plantas.
- Experimento 3:     35 fecundaciones en 10 plantas.
- Experimento 4:     40 fecundaciones en 10 plantas.
- Experimento 5:     23 fecundaciones en 5 plantas.
- Experimento 6:     34 fecundaciones en 10 plantas.
- Experimento 7:     37 fecundaciones en 10 plantas.

De un cierto número de plantas de la misma variedad solo se escogieron para la fecundación las más vigorosas. Los especímenes débiles siempre dan resultados inciertos, porque incluso en la primera generación de los híbridos, y más aún en las generaciones siguientes, algunos de los descendientes o bien no logran florecer o bien producen unas pocas semillas de calidad inferior.

---

[1] Una variedad tiene un color de vaina marrón rojizo que se transforma en violeta y azul en el momento de la maduración. El experimento con este carácter se inició durante el último año.

Además, en todos los experimentos se realizaron cruzamientos entre sí de esta manera: la variedad que, en un conjunto de fecundaciones, sirvió como portadora de las semillas, en la otra serie se empleó como planta portadora del polen.

Las plantas se desarrollaron en suelos de jardín, unas pocas también en macetas, y se mantuvieron en su posición erguida normal mediante bastones, ramas de árboles y cordeles tendidos entre ellas. Para cada experimento, se colocaron varias plantas en macetas en un invernadero durante el período de floración. Sirvieron como plantas de control para el experimento principal a cielo abierto en caso de posible perturbación por insectos. Entre los insectos que visitan la planta de guisantes, la especie de escarabajo *Bruchus pisi* podría ser perjudicial para los experimentos si aparece en un número elevado. Se sabe que la hembra de esta especie pone los huevos en la flor y, al hacer esto, abre la quilla. En los tarsos de un espécimen atrapado en una flor, se observaron varios granos de polen a través de una lupa de mano. Debe mencionarse otra circunstancia que posiblemente podría dar lugar a la introducción de polen procedente de otro individuo. En algunos casos raros, ciertas partes de una flor se atrofian, lo que provoca una exposición parcial de los órganos reproductores. Asimismo, se observó un desarrollo defectuoso de la quilla, que ocasionó que el estigma y las anteras permanecieran parcialmente descubiertos. También sucede, a veces, que el polen no alcanza su perfecto funcionamiento. En este caso, se produce un alargamiento gradual del pistilo durante la floración, hasta que el extremo del estigma sale junto a la quilla. Este notable aspecto se ha observado en híbridos de *Phaseolus* y *Lathyrus*.

El riesgo de adulteración por polen de otro individuo es muy leve para *Pisum*, y no puede perturbar el resultado general. Entre las más de 10000 plantas cuidadosamente examinadas, solo hubo unos pocos casos en los que tal interferencia ocurrió indudablemente. Puesto que en el invernadero no se encontró nunca ni un solo caso, parece acertado suponer que *Bruchus pisi* y, posiblemente, también las anomalías en la estructura floral que se han descrito fueron la causa.

**4. La Forma de los Híbridos**

Los experimentos realizados con plantas ornamentales en los años previos ya arrojaron evidencia de que los híbridos, por regla general, no son exactamente la forma intermedia entre los padres originales. Con caracteres que son particularmente notables, como los relacionados con la forma y el tamaño de las hojas y con la pubescencia de varias partes, la forma intermedia es, casi siempre, evidente. En otros casos, sin embargo, uno de los dos caracteres de los padres originales posee una dominancia tan elevada que es difícil, o casi imposible, detectar el otro en el híbrido.

Esto es precisamente lo que ocurre en los híbridos de *Pisum*. Cada uno de los siete caracteres híbridos se parece tanto a uno de los dos caracteres parentales originales que el otro, o bien escapa totalmente a la observación, o bien es tan parecido que no puede detectarse con seguridad. Esta circunstancia es de gran importancia para determinar y clasificar las formas que aparecen en la progenie de los híbridos. En este trabajo, aquellos caracteres que se transmiten completos o casi sin cambios en la asociación híbrida, y constituyen por tanto los caracteres del híbrido, se denominan *dominantes*; y aquellos caracteres que quedan latentes en el proceso se denominan *recesivos*. Se ha elegido el término recesivo porque los caracteres así designados se retiran o desaparecen por completo en el híbrido, pero reaparecen sin cambios en su progenie, como se demostrará más adelante.

Además, en todos los experimentos se demostró que no tiene ninguna importancia si el carácter dominante pertenece a la planta progenitora portadora de las semillas o portadora del polen. La forma híbrida sigue siendo exactamente la misma en ambos casos. Gärtner también se percató de este interesante hecho, y señaló que ni siquiera el experto más hábil sería capaz de determinar, en un híbrido, cuál de las dos especies progenitoras fue la portadora de las semillas o del polen.

De los caracteres diferenciales que se emplearon en los experimentos, los siguientes son dominantes:
1. La forma esférica o redondeada de la semilla, con o sin depresiones superficiales.
2. La coloración amarilla del albumen de la semilla.
3. El color gris, marrón grisáceo o marrón cuero de la cubierta de la semilla, asociado con flores de color rojo violeta y manchas rojizas en las axilas de las hojas.
4. La forma hinchada de la vaina.
5. La coloración verde de la vaina no madura, con el mismo color en el tallo, las nervaduras de las hojas y los sépalos.
6. La distribución de las flores a lo largo del tallo.
7. La mayor longitud del tallo.

Con respecto a este último carácter, hay que señalar que el tallo del híbrido generalmente supera al más largo de los dos tallos parentales, lo que puede atribuirse a la mayor exuberancia que aparece en todos los órganos de la planta cuando tallos de muy distinta longitud se unen en el híbrido. Así, por ejemplo, en repetidos experimentos, tallos de 1 pie [30,5 cm] y 6 pies [1,8 m] de largo produjeron, sin excepción, híbridos cuyos tallos variaban entre los 7 pies [2,1 m] y los 7,5 pies [2,3 m] de longitud.

En los híbridos, la cubierta de la semilla a menudo tiene más manchas, y las manchas a veces se mezclan en pequeñas manchas de color azul violeta. Las manchas a menudo aparecen incluso cuando están ausentes en los caracteres originales de los progenitores.

La forma de las semillas y el color del albumen en las formas híbridas se desarrollan inmediatamente después de la fecundación artificial, por la mera acción del polen de otro individuo. Por tanto, pueden observarse incluso en el primer año del experimento, mientras que todos los demás caracteres solo aparecen al año siguiente, en las plantas que se han desarrollado a partir de las semillas obtenidas por medio de estas fecundaciones.

## 5. La Primera Generación de los Híbridos

En esta generación reaparecen, junto con los caracteres dominantes, los caracteres recesivos con sus peculiaridades plenamente desarrolladas, cuya ocurrencia viene determinada con una proporción media de 3:1. Es decir, de cada cuatro plantas de esta generación, tres presentan el carácter dominante y una el carácter recesivo. Esto se aplica, sin excepción, a todos los caracteres que se investigaron en los experimentos. La forma angulosa y rugosa de las semillas, el color verde del albumen, el color blanco de la cubierta de las semillas y de las flores, la constricción de las vainas, el color amarillo de las vainas no maduras y de los tallos, sépalos y nervaduras de las hojas, la inflorescencia en forma de umbela y los tallos enanos reaparecen todos en la proporción numérica dada, sin alteraciones esenciales. En ninguno de los experimentos se observaron formas de transición.

Ya que los híbridos, producidos a partir de cruces entre sí, adquirieron una forma totalmente similar y no apareció ninguna variación apreciable en su desarrollo posterior, los resultados de cada experimento pudieron combinarse. Las proporciones que se obtuvieron para cada par de dos caracteres diferenciales son las siguientes:

- Experimento 1: Forma de las semillas. De 253 híbridos se obtuvieron 7324 semillas en el segundo año del experimento. De estas semillas, 5474 eran redondas o algo redondeadas, y 1850 eran angulosas y rugosas. La proporción resultante es 2,96:1.
- Experimento 2: Color del albumen. Un total de 258 plantas produjeron 8023 semillas, 6022 amarillas y 2001 verdes. Las primeras se relacionan con las segundas en la proporción 3,01:1.

En estos dos experimentos, cada vaina produjo, por lo regular, dos clases de semillas. Para las vainas bien desarrolladas, que en promedio incluían de seis a nueve semillas, a menudo todas las semillas eran redondas (primer experimento) o todas eran amarillas (segundo experimento). Sin embargo, nunca se observaron más de cinco semillas angulosas o más de cinco semillas verdes en una vaina. No parece tener importancia que la vaina se desarrolle antes o después en la planta híbrida, ni que pertenezca al tallo principal o a uno lateral. En algunas plantas, se desarrollaron pocas semillas en las vainas que se formaron primero, y estas tenían exclusivamente uno de los dos caracteres. Sin embargo, en las vainas que se formaron más tarde, se conservaron las proporciones normales. Al igual que en las diferentes vainas, la distribución de caracteres varió de manera similar entre las diversas plantas. Los primeros diez miembros de ambos conjuntos experimentales sirven como ilustración:

| Planta | Experimento 1: forma de las semillas | | Experimento 2: color del albumen | |
|---|---|---|---|---|
| | Redonda | Angulosa | Amarillo | Verde |
| 1 | 45 | 12 | 25 | 11 |
| 2 | 27 | 8 | 32 | 7 |
| 3 | 24 | 7 | 14 | 5 |
| 4 | 19 | 10 | 70 | 27 |
| 5 | 32 | 11 | 24 | 13 |
| 6 | 26 | 6 | 20 | 6 |
| 7 | 88 | 24 | 32 | 13 |
| 8 | 22 | 10 | 44 | 9 |
| 9 | 28 | 6 | 50 | 14 |
| 10 | 25 | 7 | 44 | 18 |

Como extremos en la distribución de los dos caracteres de semillas observados en una planta, en el primer experimento, en un caso, 43 semillas eran redondas y solo dos angulosas; y en otro caso, 14 eran redondas y 15 angulosas. En el segundo experimento, hubo un caso con 32 semillas amarillas y solo una verde; y en otro caso, 20 semillas eran amarillas y 16 verdes.

Estos dos experimentos son importantes para determinar las proporciones medias, porque demuestran que, si el número de plantas experimentales es menor, pueden ocurrir muy considerables oscilaciones. Al contar las semillas, especialmente en el segundo experimento, se requiere algún cuidado, ya que, en algunas de las semillas de muchas plantas, el color verde del albumen está menos desarrollado y, al principio, puede pasar fácilmente desapercibido. Esta desaparición parcial de la coloración verde no tiene conexión con el carácter híbrido de las plantas porque ocurre también en la planta progenitora. Este fenómeno de la decoloración se limita además al individuo y no la hereda la progenie. En las plantas exuberantes se observó frecuentemente este rasgo. Las semillas dañadas por insectos durante su desarrollo varían a menudo de color y de forma, pero, con un poco de práctica para clasificarlas, se evitan fácilmente los errores. Es casi superfluo mencionar que las vainas deben permanecer en la planta hasta que hayan madurado por completo y se hayan secado, porque solo entonces la forma y el color de las semillas están completamente desarrollados.

- Experimento 3: Color de la cubierta de las semillas. De 929 plantas, 705 produjeron flores de color rojo violeta y cubiertas de semillas de color marrón grisáceo; 224 tenían flores blancas y cubiertas de las semillas blancas. Esto da como resultado una proporción 3,15:1.
- Experimento 4: Forma de las vainas. De 1181 plantas, 882 tenían las vainas hinchadas, y 299 tenían vainas con constricciones. Por tanto, la relación es 2,95:1.
- Experimento 5: Color de la vaina no madura. El número de plantas experimentales fue de 580, de las cuales 428 tenían vainas verdes, y 152 tenían vainas amarillas. Por tanto, están en la proporción 2,82:1.
- Experimento 6: Posición de las flores. De 858 casos, las flores se ubicaron a lo largo del tallo 651 veces, y fueron terminales 207 veces. La proporción es 3,14:1.
- Experimento 7: Longitud del tallo. De 1064 plantas, 787 tenían tallos largos, y 277 tenían tallos cortos. Por tanto, la proporción es 2,84:1. En este experimento, las plantas enanas fueron cuidadosamente trasplantadas a un suelo especial. Esta precaución era necesaria porque, de otro modo, hubieran perecido por quedar cubiertas por sus parientes altos. Incluso en sus fases más juveniles pueden reconocerse fácilmente por su porte compacto y por su follaje verde oscuro.

Si se combinan los resultados de todos los experimentos, se obtiene una relación promedio entre el número de formas con caracteres dominantes y recesivos de 2,98:1, esto es, 3:1.

El carácter dominante puede tener aquí una doble significación, a saber, la de un carácter parental original o la de un carácter híbrido. Cuál de estos dos significados ocurre en cada caso solo puede determinarse en la próxima generación. Un carácter parental original debe transmitirse sin cambios a toda la progenie, mientras que el carácter híbrido debe seguir el mismo comportamiento que se observó en la primera generación.

## 6. La Segunda Generación de los Híbridos

Aquellas formas que conservan el carácter recesivo en la primera generación no varían en la segunda generación en relación con ese carácter. Permanecen constantes en su progenie.

Este no es el caso de aquellas formas que poseen el carácter dominante en la primera generación. De ellas, dos tercios dan progenie que porta el carácter dominante y recesivo en una proporción 3:1. Por tanto, muestran el mismo comportamiento que las formas híbridas, mientras que solo un tercio permanece constante con el carácter dominante.

Los distintos experimentos produjeron los siguientes resultados:

- Experimento 1: De las 565 plantas cultivadas a partir de semillas redondas de la primera generación, 193 produjeron únicamente semillas redondas y, por tanto, permanecieron constantes en este carácter. Sin embargo, 372 plantas produjeron simultáneamente semillas redondas y angulosas en una proporción de 3:1. Consecuentemente, el número de formas híbridas en relación con el número de formas constantes fue de 1,93:1.
- Experimento 2: De las 519 plantas cultivadas a partir de semillas cuyo albumen en la primera generación tenía color amarillo, 166 produjeron exclusivamente albumen amarillo, mientras que 353 plantas produjeron semillas con albumen amarillo y verde en una proporción 3:1. Esto dio como resultado una separación entre formas híbridas y formas constantes en una proporción 2,13:1.

Para cada uno de los siguientes experimentos, se seleccionaron 100 plantas que conservaron el carácter dominante en la primera generación y, para probar su significado, se cultivaron diez semillas de cada una de ellas:

- Experimento 3: La progenie de 36 plantas produjo exclusivamente semillas con cubierta de color marrón grisáceo. Sin embargo, en la progenie de 64 plantas, algunas semillas tenían una cubierta de color marrón grisáceo, y otras, una cubierta blanca.
- Experimento 4: La progenie de 29 plantas solo tenía vainas hinchadas. Sin embargo, en la progenie de 71 plantas, algunas tenían vainas hinchadas, y otras, vainas con constricciones.
- Experimento 5: La progenie de 40 plantas tenía solo vainas verdes. Sin embargo, en la progenie de 60 plantas, algunas tenían vainas verdes, y otras, vainas amarillas.
- Experimento 6: La progenie de 33 plantas tenía flores ubicadas solo a lo largo del tallo. Sin embargo, en la progenie de 67 plantas, algunas tenían flores ubicadas a lo largo del tallo, y otras, flores terminales.
- Experimento 7: La progenie de 28 plantas produjo tallos largos. Sin embargo, en la progenie de 72 plantas, algunas tenían tallos largos, y otras, tallos cortos.

En cada uno de estos experimentos es constante un número particular de plantas con el carácter dominante. Para determinar la proporción en que se presentan las formas con carácter constantemente persistente, tienen especial importancia los dos primeros experimentos, ya que en ellos puede compararse un mayor número de plantas. Las proporciones 1,93:1 y 2,13:1, tomadas en conjunto, dan como resultado, casi con exactitud, la proporción media 2:1. El sexto experimento tiene un resultado casi idéntico. En los demás experimentos, la proporción fluctúa más o menos, como cabía esperar, dado el pequeño número de 100 plantas experimentales. El quinto experimento, que mostró la mayor desviación, se repitió y, como resultado, en lugar de la proporción 60:40, produjo la relación 65:35. Por tanto, la proporción media 2:1 parece cierta. En consecuencia, se demuestra que, de las formas que poseen el carácter dominante en la primera generación, dos tercios tienen el carácter híbrido, mientras que un tercio permanece constantemente con el carácter dominante.

Así pues, la proporción 3:1 gobierna la distribución de los caracteres dominantes y recesivos en la primera generación. Ahora bien, si se distingue simultáneamente el carácter dominante en su significado como carácter híbrido y como carácter parental original, la proporción anterior se convierte, consecuentemente, en la proporción 2:1:1 para todos los experimentos. Debido a que los miembros de la primera generación proceden directamente de las semillas de los híbridos, se hace evidente ahora que los híbridos producen semillas de cada par de caracteres diferenciales. De estas semillas, la mitad vuelve a desarrollar la forma híbrida, mientras que la otra mitad produce plantas que permanecen constantes y reciben, a partes iguales, el carácter dominante o el recesivo.

### 7. Las Generaciones Subsiguientes de los Híbridos

Las proporciones en que se desarrollan y segregan los descendientes de los híbridos en la primera y segunda generación son válidas, con toda probabilidad, para todas las generaciones subsiguientes. El primer y el segundo experimento se han llevado a cabo durante seis generaciones; el tercer y el séptimo experimento, durante cinco generaciones; y el cuarto, quinto y sexto experimento, durante cuatro generaciones. Ha de advertirse que estos experimentos, a partir de la tercera generación, se han realizado con un número menor de plantas, pero no se han encontrado desviaciones perceptibles de la regla. La progenie de los híbridos, en cada generación, se segregó en formas híbridas y constantes según la proporción 2:1:1.

Si *A* representa uno de los dos caracteres constantes, por ejemplo, el dominante, *a* representa el recesivo, y *Aa* representa la forma híbrida en la que los dos caracteres están unidos, entonces la serie de desarrollo

$$A + 2Aa + a$$

representa la progenie de los híbridos de cada par de caracteres diferenciales.

Las observaciones realizadas por Gärtner, Kölreuter y otros, con respecto al hecho de que los híbridos tienen la tendencia de volver a la formas parentales originales, se confirman en los experimentos que se describen aquí. Se demuestra que el número de híbridos descendientes de la fecundación disminuye significativamente de generación en generación pero sin desaparecer por completo, en comparación con el número de formas y su progenie que se han vuelto constantes. Si suponemos que, en promedio, todas las plantas de todas las generaciones tienen una fertilidad media igual, y si consideramos además que cada híbrido forma semillas de las cuales la mitad surgen nuevamente como híbridos, mientras que la otra mitad se vuelve constante para ambos caracteres en partes iguales, entonces las proporciones de la progenie en cada generación se pueden mostrar mediante la siguiente tabulación, donde *A* y *a* nuevamente representan los dos caracteres originales, y *Aa* representa la forma híbrida. Para ser breve, puede suponerse que cada planta, en cada generación, solo produce cuatro semillas.

| Generación | *A* | *Aa* | *a* | Proporciones A : Aa : a |
|---|---|---|---|---|
| 1 | 1 | 2 | 1 | 1 : 2 : 1 |
| 2 | 6 | 4 | 6 | 3 : 2 : 3 |
| 3 | 28 | 8 | 28 | 7 : 2 : 7 |
| 4 | 120 | 16 | 120 | 15 : 2 : 15 |
| 5 | 496 | 32 | 496 | 31 : 2 : 31 |
| n | | | | $2^n - 1$ : 2 : $2^n - 1$ |

En la décima generación, por ejemplo, hay $2^n - 1 = 1023$. De cada 2048 plantas que se originan en esta generación, hay 1023 que tienen el carácter dominante constante, otras 1023 con el carácter recesivo constante, y solo dos híbridos.

**8. La Progenie de los Híbridos en los que se Combinan Varios Caracteres Diferenciales**

En los experimentos descritos previamente, se emplearon plantas que diferían en un solo carácter esencial. El siguiente objetivo consistió en investigar si la ley del desarrollo encontrada para cada par de caracteres diferenciales era válida cuando varios caracteres diversos se unen en el híbrido por fecundación.

En cuanto a la forma de los híbridos en este caso, los experimentos demostraron siempre que el híbrido se aproxima más a la planta progenitora original que posee el mayor número de caracteres dominantes. Si, por ejemplo, la planta productora de las semillas tiene el tallo corto, flores blancas terminales y vainas hinchadas, mientras que la planta productora de polen tiene el tallo largo, flores de color rojo violeta a lo largo del tallo y vainas con constricciones, entonces, el híbrido se parece al progenitor productor de semillas solamente en la forma de la vaina; en los demás caracteres, el híbrido es idéntico al progenitor productor de polen. Si uno de los dos progenitores originales poseyera únicamente caracteres dominantes, entonces, el híbrido apenas se distinguiría de él.

Se realizaron dos experimentos con un número considerable de plantas. En el primer experimento, las plantas progenitoras originales diferían en la forma de las semillas y en el color del albumen. En el segundo experimento, las plantas progenitoras originales diferían en la forma de las semillas, en el color del albumen y en el color de la cubierta de las semillas. Los experimentos con los caracteres de las semillas dan resultados más sencillos y más seguros.

Para facilitar el estudio de los datos de estos experimentos, los diferentes caracteres de la planta productora de semillas se designan con *A*, *B* y *C*; los diferentes caracteres de la planta productora de polen, con *a*, *b*, y *c*; y las formas híbridas de estos caracteres, con *Aa*, *Bb* y *Cc*:

**Experimento 1**

| *AB*, progenitores productores de la semilla | *ab*, progenitores productores de polen |
|---|---|
| *A*, forma redonda | *a*, forma rugosa |
| *B*, albumen amarillo | *b*, albumen verde |

Las semillas fecundadas resultaron redondas y amarillas, como las de las plantas progenitoras productoras de semillas. Las plantas que se desarrollaron luego produjeron semillas de cuatro clases, que, a menudo, se presentaban en una misma vaina. En total se obtuvieron 556 semillas de 15 plantas, y en esta muestra había: 315 semillas redondas y amarillas, 101 rugosas y amarillas, 108 redondas y verdes, y 32 rugosas y verdes.

Todas estas semillas se sembraron al año siguiente. De las semillas redondas y amarillas, 11 no produjeron plantas, y tres plantas no produjeron semillas. Entre las plantas restantes, había 38 semillas redondas y amarillas, *AB*; 65 semillas redondas amarillas y verdes, *ABb*; 60 semillas redondas amarillas y rugosas amarillas, *AaB*; y 138 semillas redondas amarillas y verdes, rugosas amarillas y verdes, *AaBb*.

De las semillas rugosas amarillas, 96 plantas produjeron semillas, de las cuales: 28 tenían solamente semillas rugosas amarillas, *aB*; y 68 tenían semillas rugosas amarillas y verdes, *aBb*.

De las 108 semillas redondas verdes, 102 produjeron plantas que fructificaron, de las cuales: 35 solo tenían semillas redondas y verdes, *Ab*; y 67 tenían semillas redondas, verdes y rugosas, *Aab*.

Las semillas rugosas y verdes dieron lugar a 30 plantas que produjeron todas las semillas con los mismos caracteres y permanecieron constantes, *ab*.

En la progenie de los híbridos aparecieron, por tanto, 9 formas diferentes, algunas de ellas en números muy desiguales. Cuando estas se recopilaron y se ordenaron, se obtuvieron los siguientes resultados:

| | | | |
|---|---|---|---|
| 38 | plantas | designadas con | *AB* |
| 35 | plantas | designadas con | *Ab* |
| 28 | plantas | designadas con | *aB* |
| 30 | plantas | designadas con | *ab* |
| 65 | plantas | designadas con | *ABb* |
| 68 | plantas | designadas con | *aBb* |
| 60 | plantas | designadas con | *AaB* |
| 67 | plantas | designadas con | *Aab* |
| 138 | plantas | designadas con | *AaBb* |

El conjunto de formas puede clasificarse en tres grupos, esencialmente diferentes:
- El primero incluye aquellas formas con las designaciones *AB*, *Ab*, *aB* y *ab*. Poseen solo caracteres constantes y no vuelven a variar en las generaciones posteriores. Cada una de estas formas aparece, en promedio, 33 veces.
- El segundo grupo incluye las formas *ABb*, *aBb*, *AaB* y *Aab*. Estas formas son constantes en un carácter, híbridos en el otro, y en la próxima generación solo varían en el carácter híbrido. Cada una de estas formas aparece, en promedio, 65 veces.
- La forma *AaBb* aparece 138 veces, es híbrida en ambos caracteres y se comporta exactamente como los híbrido de los que procede.

Si se comparan los números en que aparecen las formas pertenecientes a estas clases, son evidentes las proporciones 1:2:4. Los números 38, 65, 138 se aproximan mucho a los números de la proporción 33, 66, 132.

La serie de desarrollo consta de nueve clases. Cuatro de ellas aparecen solo una vez cada una y son constantes en los dos caracteres. Las formas *AB* y *ab* se parecen a los progenitores originales. Las otras dos presentan combinaciones entre las uniones de los caracteres *A*, *a*, *B*, *b*; estas combinaciones también es posible que sean constantes. Cuatro clases aparecen dos veces cada una, son constantes en un carácter, e híbridas en el otro. Una clase ocurre cuatro veces y es híbrida en ambos caracteres. Por consiguiente, la progenie de los híbridos, cuando se combinan en ellos dos pares de caracteres diferentes, se desarrolla según la siguiente serie:

$$AB + Ab + aB + ab + 2ABb + 2aBb + 2AaB + 2Aab + 4AaBb$$

No hay duda de que es una serie combinatoria en la que las dos series de desarrollo de los caracteres *A* y *a*, *B* y *b*, están combinadas. Llegamos al número completo de clases en la serie cuando se combinan los términos:

$$A + 2Aa + a$$
$$B + 2Bb + b$$

**Experimento 2**

| *ABC*, plantas productoras de semillas | *abc*, plantas productoras de polen |
|---|---|
| *A*, forma redonda | *a*, forma rugosa |
| *B*, albumen amarillo | *b*, albumen verde |
| *C*, cubierta de la semilla marrón grisácea | *c*, cubierta de la semilla blanca |

Este experimento se llevó a cabo de manera bastante similar al anterior. De todos los experimentos, este requirió más tiempo y cuidados. De 24 híbridos, se obtuvieron un total de 687 semillas, todas manchadas, de color marrón grisáceo o verde grisáceo, redondas o angulosas. A partir de estas, 639 plantas produjeron semillas al año siguiente y, como demostraron investigaciones posteriores, entre ellas se encontraban las siguientes formas:

| | | | | | | | | | | | |
|---|---|---|---|---|---|---|---|---|---|---|---|
| 8 | plantas | *ABC* | 22 | plantas | *ABCc* | 45 | plantas | *ABbCc* | 78 | plantas | *AaBbCc* |
| 14 | plantas | *ABc* | 17 | plantas | *AbCc* | 36 | plantas | *aBbCc* | | | |
| 9 | plantas | *AbC* | 25 | plantas | *aBCc* | 38 | plantas | *AaBCc* | | | |
| 11 | plantas | *Abc* | 20 | plantas | *abCc* | 40 | plantas | *AabCc* | | | |
| 8 | plantas | *aBC* | 15 | plantas | *ABbC* | 49 | plantas | *AaBbC* | | | |
| 10 | plantas | *aBc* | 18 | plantas | *ABbc* | 48 | plantas | *AaBbc* | | | |
| 10 | plantas | *abC* | 19 | plantas | *aBbC* | | | | | | |
| 7 | plantas | *abc* | 24 | plantas | *aBbc* | | | | | | |
| | | | 14 | plantas | *AaBC* | | | | | | |
| | | | 18 | plantas | *AaBc* | | | | | | |
| | | | 20 | plantas | *AabC* | | | | | | |
| | | | 16 | plantas | *Aabc* | | | | | | |

La serie de desarrollo incluye 27 clases, en las que: ocho tienen los caracteres constantes, y cada uno aparece, en promedio, diez veces; 12 son constantes en dos caracteres, híbridos en el tercero, y cada uno aparece, en promedio, 19 veces; seis son constantes en un carácter, híbridos en los otros dos, y cada uno de ellos ocurre, en promedio, 43 veces; y una forma aparece 78 veces, y es híbrida para todos los caracteres. La proporción 10:19:43:78 concuerda tan estrechamente con la proporción 10:20:40:80, esto es, la proporción 1:2:4:8, que esta última, sin duda, representa los verdaderos valores.

El desarrollo de los híbridos, cuando sus progenitores originales difieren en tres caracteres, viene gobernado por la siguiente serie de desarrollo:

*ABC* + *ABc* + *AbC* + *Abc* + *aBC* + *aBc* + *abC* + *abc* + *2ABCc* + *2AbCc* + *2aBCc* + *2abCc* + *2ABbC* + *2ABbc* + *2aBbC* + *2aBbc* + *2AaBC* + *2AaBc* + *2AabC* + *2Aabc* + *4ABbCc* + *4aBbCc* + *4AaBCc* + *4AabCc* + *4AaBbC* + *4AaBCc* + *8AaBbCc*

En este caso también se trata de una serie combinatoria, en la que las series de desarrollo de los caracteres *A* y *a*, *B* y *b*, *C* y *c* están asociadas entre sí. Los términos:

$$A + 2Aa + a$$
$$B + 2Bb + b$$
$$C + 2Cc + c$$

reflejan todas las clases de la serie. Se presentan las combinaciones constantes que son posibles entre los caracteres *A*, *B*, *C*, *a*, *b*, *c*. Dos de esa combinaciones, *ABC* y *abc*, se asemejan a las dos plantas parentales originales.

Además, se realizaron otros experimentos con un número menor de plantas experimentales. En estos experimentos, los restantes caracteres se unieron de dos en dos y de tres en tres en los híbridos. Todos produjeron aproximadamente los mismos resultados. Consecuentemente, no hay duda de que, a todos los caracteres que intervinieron en los experimentos, se les aplica el siguiente principio: La progenie de los híbridos, en los que se unen varios caracteres esenciales diferentes, presenta los términos de una serie de combinaciones, en las que la serie de desarrollo para cada par de caracteres diferenciales se combinan. Asimismo, se demuestra que el comportamiento de cada par de caracteres diferentes en la unión híbrida es independiente de otras diferencias entre las dos plantas parentales originales.

Si $n$ representa el número de caracteres diferenciales en las dos plantas progenitoras originales, entonces, $3^n$ da el número de términos en la serie de combinaciones, $4^n$ proporciona el número de individuos que pertenecen a la serie, y $2^n$ ofrece el número de uniones que permanecen constantes. Así, por ejemplo, si los progenitores originales difieren en 4 caracteres, la serie incluye $3^4 = 81$ clases, $4^4 = 256$ individuos, y $2^4 = 16$ formas constantes. Dicho de otro modo, de cada 256 descendientes de los híbridos, hay 81 combinaciones diferentes, de las cuales 16 son constantes.

Todas las combinaciones constantes que son posibles en *Pisum*, mediante la combinación de los siete caracteres diferenciales indicados, se obtuvieron realmente mediante cruzamientos repetidos. Su número resulta del cálculo $2^7 = 128$. Esto constituye, por tanto, una prueba de que los caracteres constantes que aparecen en las diferentes formas de un género de plantas pueden ocurrir, a través de la fecundación artificial repetida, en todas las combinaciones posibles de acuerdo con las leyes matemáticas de la combinatoria.

Los experimentos sobre el tiempo de floración de los híbridos todavía no han concluido. No obstante, ya es posible señalar, a este respecto, que la floración tiene lugar en un momento casi exactamente intermedio entre la floración del progenitor productor de la semilla y la floración de progenitor productor del polen. El desarrollo de los híbridos en relación con este carácter seguirá, probablemente, la regla indicada en el caso de los demás caracteres. Las formas elegidas para experimentos de esta clase deben diferir, en el tiempo medio de floración, en al menos 20 días. Además, es necesario que las semillas, al sembrarse, se coloquen a la misma profundidad en la tierra para que germinen simultáneamente. Asimismo, durante todo el periodo de floración, se deben tener en cuenta las grandes fluctuaciones de temperatura, porque pueden explicar que la floración se adelante o se retrase parcialmente. Es evidente que estos experimentos presentan muchas dificultades y que necesitan gran atención.

En síntesis, los resultados conseguidos indican que aquellos caracteres diferenciales que admiten un reconocimiento fácil y seguro en las plantas experimentales se comportan todos exactamente igual en la unión híbrida. La mitad de la progenie de los híbridos, para cada par de caracteres diferenciales, también es híbrida, mientras que la otra mitad es constante y exhibe igual proporción con los caracteres de la planta productora de semillas que con los de la productora del polen, respectivamente. Si varios caracteres diferenciales se unen en un híbrido a través de la fecundación, la progenie de este híbrido constituye los términos de una serie combinatoria, que resulta de la unión de las series de desarrollo de todos los pares de caracteres diferenciales.

El comportamiento uniforme que muestra el conjunto de caracteres investigados permite, y justifica plenamente, la aceptación del principio de que existe una relación semejante en otros caracteres que aparecen menos claramente definidos en las plantas y que, por tanto, no pudieron incluirse en los diversos experimentos. Un experimento con pedúnculos de longitudes diferentes dio, en conjunto, resultados muy satisfactorios, a pesar de que la diferenciación y disposición de las formas no pudiera realizarse con la seguridad y rigurosidad que se requerían para que el experimento fuera correcto.

**9. Las Células Reproductoras de los Híbridos**

Los resultados de los experimentos descritos previamente condujeron a otros experimentos cuyos resultados parecen apropiados para extraer algunas conclusiones sobre la constitución de las células huevo y las células del polen de los híbridos. *Pisum* supone un excelente punto de partida por dos motivos: (1) aparecen formas constantes en la progenie de los híbridos; y (2) también aparecen formas constantes para todas las combinaciones de los caracteres asociados. A juzgar por los resultados de los experimentos, se confirma que, para todos los casos, la progenie constante puede formarse solo cuando la célula huevo y la del polen fecundante tengan los mismos caracteres, es decir, cuando ambas células tengan la capacidad de crear individuos completamente similares, como ocurre en la fecundación normal en las especies puras. Por consiguiente, debemos aceptar como cierto que, en la producción de formas constantes en las plantas híbridas, también deben actuar factores exactamente iguales. Puesto que una planta, o incluso una flor de una planta, produce diferentes formas constantes, parece razonable inferir que: (1) en los ovarios de los híbridos se crean tantas clases de células huevo, y en las anteras se originan tantas clases de células del polen, como formas de combinaciones *constantes* son posibles; y (2) estas células huevo y del polen tienen una composición interna que concuerda con la de las formas individuales.

De hecho, puede demostrarse teóricamente que esta hipótesis sería completamente suficiente para explicar el desarrollo de los híbridos en cada una de las generaciones si, además, se supone que, en los híbridos, las diferentes clases de células huevo y de polen se crean, en promedio, en igual número.

Para probar estas hipótesis de investigación, se diseñó el siguiente experimento: dos tipos que eran constantemente diferentes en la forma de la semilla y en el color del albumen se unieron por fecundación. Si los caracteres diferenciales se representan, de nuevo, como *A*, *B*, *a*, *b*, se tiene:

| *AB*, progenitor productor de semilla | *ab*, progenitor productor de polen |
|---|---|
| *A*, forma redonda | *a*, forma angulosa |
| *B*, albumen amarillo | *b*, albumen verde |

Las semillas fecundadas artificialmente se sembraron junto con semillas de las dos plantas originales, y se seleccionaron los especímenes más vigorosos para cruzamientos entre sí. Se fecundaron:

1. Los híbridos con el polen *AB*.
2. Los híbridos con el polen *ab*.
3. *AB* con el polen de los híbridos.
4. *ab* con el polen de los híbridos.

En cada uno de estos cuatro experimentos, se fecundaron todas las flores de tres plantas. Si la suposición anterior es correcta, entonces, las células huevo y las células de polen deberían desarrollarse en el híbrido como formas *AB*, *Ab*, *aB*, *ab*, cuyas combinaciones deberían ser las siguientes:

1. Las células huevo *AB*, *Ab*, *aB*, *ab* con las célula de polen *AB*.
2. Las células huevo *AB*, *Ab*, *aB*, *ab* con las células de polen *ab*.
3. Las células huevo *AB* con las células de polen *AB*, *Ab*, *aB*, *ab*.
4. Las células huevo *ab* con las células de polen *AB*, *Ab*, *aB*, *ab*.

Con estas condiciones, de cada uno de los experimentos podrían surgir únicamente las siguientes formas:

1. *AB*, *ABb*, *AaB*, *AaBb*.
2. *AaBb*, *Aab*, *aBb*, *ab*.
3. *AB*, *ABb*, *AaB*, *AaBb*.
4. *AaBb*, *Aab*, *aBb*, *ab*.

Además, si las diferentes formas de las células huevo y de polen de los híbridos se crearan, en promedio, en igual número, entonces, en cada experimento, las cuatro combinaciones indicadas anteriormente deberían estar en la misma proporción entre sí. Sin embargo, no era de esperar una concordancia perfecta en las proporciones numéricas porque, en cada fecundación, incluidas las fecundaciones que no son artificiales, algunas células huevo no se desarrollaron o se atrofiaron más tarde, e incluso algunas de las semillas bien desarrolladas no germinaron tras la siembra. La suposición anterior se haya limitada también por el hecho de que, si bien requiere que se creen las diferentes formas de células huevo y de polen en igual número, no es necesario que esto se cumpla en cada híbrido con exactitud matemática.

Los experimentos primero y segundo tenían como objetivo principal verificar la composición de las células huevo de los híbridos; y los experimentos tercero y cuarto, determinar la composición de las células de polen. Como muestra la tabla anterior, los experimentos primero y tercero deberían producir las mismas combinaciones que los experimentos segundo y cuarto. Incluso ya en el segundo año, el resultado debería ser parcialmente visible en la forma y en el color de las semillas fecundadas artificialmente.

En los experimentos primero y tercero, los caracteres dominantes de la forma y del color, *A* y *B*, aparecen en cada combinación, una parte en asociación constante y la otra parte en unión híbrida con los caracteres recesivos *a* y *b*, por lo que deben imprimir sus peculiaridades en todas las semillas. Por tanto, si la suposición es cierta, todas las semillas deberían ser redondas y amarillas.

En cambio, en los experimentos segundo y cuarto: (1) una combinación es híbrida en la forma y en el color, consecuentemente, las semillas deberían ser redondas y amarillas; (2) otra combinación es híbrida en la forma, pero es constante en el carácter recesivo del color, así pues, las semillas deberían ser redondas y verdes; (3) la tercera combinación es constante en el carácter recesivo de la forma, pero es híbrida en el color, por tanto, las semillas deberían ser angulosas y amarillas; y (4) la cuarta combinación es constante en ambos caracteres recesivos, de manera que las semillas deberían ser angulosas y verdes. En resumen, en estos dos experimentos se esperaban cuatro clases de semillas: redondas y amarillas, redondas y verdes, angulosas y amarillas, y angulosas y verdes.

La cosecha cumplió completamente con estas previsiones, a saber:
- En el primer experimento se obtuvieron 98 semillas exclusivamente redondas y amarillas.
- En el tercer experimento se obtuvieron 94 semillas exclusivamente redondas y amarillas.
- En el segundo experimento se obtuvieron 31 semillas redondas y amarillas, 26 redondas y verdes, 27 angulosas y amarillas, y 26 angulosas y verdes.
- En el cuarto experimento se obtuvieron 24 semillas redondas y amarillas, 25 redondas y verdes, 22 angulosas y amarillas, y 26 angulosas y verdes.

Ya no había ninguna duda sobre el éxito del experimento, pero la próxima generación proporcionaría la prueba final. Así pues, al año siguiente, de las semillas sembradas fructificaron 90 plantas en el primer experimento, y 87 plantas en el tercer experimento, con la distribución que se indica a continuación:

| 1.er Experimento | 3.er Experimento | Características de las semillas | Forma |
|---|---|---|---|
| 20 | 25 | Semillas redondas y amarillas | ***AB*** |
| *23* | 19 | Semillas redondas, amarillas o verdes | ***ABb*** |
| 25 | 22 | Semillas redondas o angulosas verdes | ***AaB*** |
| *22* | 21 | Semillas redondas o angulosas, verdes o amarillas | ***AaBb*** |

En los experimentos segundo y cuarto:
- Las semillas redondas y amarillas produjeron plantas con semillas redondas o angulosas, amarillas o verdes, ***AaBb***.
- A partir de las semillas redondas y verdes, resultaron plantas con semillas redondas o angulosas de color verde, ***Aab***.

- Las semillas angulosas y amarillas dieron plantas con semillas angulosas de color amarillo o verde, *aBb*.
- De las semillas angulosas y verdes, crecieron plantas que produjeron, una vez más, solo semillas angulosas y verdes, *ab*.

Aunque algunas semillas tampoco germinaron en estos dos experimentos, las cifras obtenidas en el año anterior no quedaron afectadas, ya que, de cada clase de semillas, surgieron plantas que, en lo que se refiere a sus semillas, eran similares entre sí y diferentes de las otras. En concreto, el recuento en estos dos experimentos se corresponde con la siguiente tabla:

| 2.º Experimento | 4.º Experimento | | Forma |
|---|---|---|---|
| 31 | 24 | plantas con semillas con la combinación | *AaBb* |
| 26 | 25 | plantas con semillas con la combinación | *Aab* |
| 27 | 22 | plantas con semillas con la combinación | *aBb* |
| 26 | 27 | plantas con semillas con la combinación | *ab* |

En todos los experimentos, por tanto, aparecieron todas las formas que demandaba nuestra suposición, y casi en igual número.

En otra prueba posterior, se experimentó con los caracteres del color de la flor y la longitud del tallo. La selección se diseñó de modo que, en el tercer año del experimento, cada carácter apareciera en la mitad de todas las plantas si la suposición anterior fuera cierta. De nuevo, si los diferentes caracteres se designan como *A*, *B*, *a*, *b*, tenemos:

| *A*, flores rojo violeta | *a*, flores blancas |
|---|---|
| *B*, tallo largo | *b*, tallo corto |

La forma *Ab* se fecundó con *ab*, que produjo el híbrido *Aab*. Además, *aB* se fecundó también con *ab*, de donde se obtuvo el híbrido *aBb*. En el segundo año, para una fecundación posterior, se empleó el híbrido *Aab* como planta productora de semillas, y el híbrido *aBb*, como planta productora de polen:

| Planta productora de la semilla, *Aab* | Planta productora de polen, *aBb* |
|---|---|
| Posibles células huevo, *Ab*, *ab* | Células del polen, *aB*, *ab* |

De la fecundación entre las posibles células huevo y de polen deberían surgir cuatro combinaciones, a saber:

$$AaBb + aBb + Aab + ab$$

De acuerdo con la suposición anterior, en el tercer año del experimento, en las plantas que se obtengan debería existir la siguiente distribución:

| | | |
|---|---|---|
| La mitad debería tener flores de color rojo violeta | *Aa* | Clases 1, 3 |
| La mitad debería tener flores blancas | *a* | Clases 2, 4 |
| La mitad debería tener un tallo largo | *Bb* | Clases 1, 2 |
| La mitad debería tener un tallo corto | *b* | Clases 3, 4 |

Así pues, de las 45 fecundaciones realizadas en el segundo año, resultaron 187 semillas, de las cuales únicamente 166 alcanzaron la fase de floración en el tercer año. En estas últimas, las diferentes clases aparecieron con la distribución que se tabula a continuación:

| Clase | Color de las flores | Tallo | Cantidad |
|---|---|---|---|
| 1 | rojo violeta | largo | 47 veces |
| 2 | blanco | largo | 40 veces |
| 3 | rojo violeta | corto | 38 veces |
| 4 | blanco | corto | 41 veces |

Y el número de plantas que presentaban cada una de las cuatro formas fue:

| Carácter | Forma | Cantidad |
|---|---|---|
| Color rojo violeta de las flores | *Aa* | 85 plantas |
| Color blanco de las flores | *a* | 81 plantas |
| Tallo largo | *Bb* | 87 plantas |
| Tallo corto | *b* | 79 plantas |

Por tanto, la teoría propuesta queda confirmada satisfactoriamente con este experimento.

Para los caracteres de la forma de la vaina, el color de la vaina y la posición de la flor, se realizaron experimentos a una menor escala, y los resultados obtenidos concuerdan plenamente. Es decir, todas las combinaciones que eran posibles por la unión de los caracteres diferenciales aparecieron como se esperaba y en números casi iguales.

Consecuentemente, queda confirmada experimentalmente la teoría de que los híbridos de guisantes forman células huevo y de polen que, en su constitución, representan en igual número todas las formas constantes que surgen de la combinación de los caracteres unidos en la fecundación.

Las diferentes formas en la progenie de los híbridos, así como las proporciones numéricas en que se observan, quedan explicadas suficientemente con el principio que se acaba de formular. El caso más simple lo ofrece la serie de desarrollo para cada par de caracteres diferenciales. Esta serie está representada por la expresión:

$$A + 2Aa + a$$

en la que *A* y *a* son las formas con caracteres diferenciales constantes, y *Aa* es la forma híbrida de ambos caracteres. La serie incluye cuatro individuos en tres clases diferentes. En la creación de dichos individuos, las células huevo y del polen de las formas *A* y *a* se dan, por término medio, en proporciones iguales en la fecundación; así, cada forma se presenta dos veces porque se originan cuatro individuos. Por tanto, en la fecundación participan:

- Las células de polen: *A* + *A* + *a* + *a*
- Las células huevo: *A* + *A* + *a* + *a*

Por consiguiente, es aleatorio cuál de las dos clases de polen se une con cada célula huevo. De acuerdo con la teoría de la probabilidad, en el promedio de muchos casos, cada forma de polen *A* y *a* se une, con la misma frecuencia, con cada forma de célula huevo *A* y *a*. Así, una de las dos células de polen *A* converge con una célula huevo *A* durante la fecundación, y la otra, con una célula huevo *a*. De la misma manera, una célula de polen *a* se une con una célula huevo *A*, y la otra, con una célula huevo *a*.

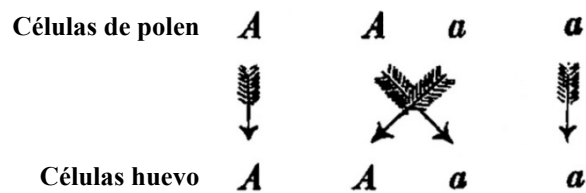

Si la unión de las células huevo con las del polen se expresa mediante líneas de fracción, de manera que la forma de la célula de polen se coloca en el numerador, y la forma de la célula huevo, en el denominador, el resultado de la fecundación se puede ilustrar con esta suma de fracciones:

$$\frac{A}{A} + \frac{A}{a} + \frac{a}{A} + \frac{a}{a}$$

En esta suma, la primera y la cuarta fracción corresponden a la unión de una célula de polen y una célula huevo de la misma forma; por tanto, el individuo que se origina de la asociación debe poseer caracteres constantes, esto es, *A* y *a*. Sin embargo, la segunda y la tercera fracción representan la unión, una vez más, de los dos caracteres diferenciales de los parientes originales; así pues, las formas que aparecen de esta fecundación son completamente idénticas al híbrido del que proceden. En consecuencia, tiene lugar una hibridación repetida. Esto explica el fenómeno remarcable de que los híbridos, al igual que las dos formas parentales originales, sean capaces de producir progenie que sea idéntica a ellos mismos. Tanto la unión *A*/*a* como la unión *a*/*A* produce la combinación de caracteres *Aa* porque, como ya se ha indicado anteriormente, no afecta al resultado de la fecundación cuál de los dos caracteres pertenezca a la célula de polen o a la célula huevo. Siendo así, podemos acudir a la siguiente expresión:

$$\frac{A}{A} + \frac{A}{a} + \frac{a}{A} + \frac{a}{a} = A + 2Aa + a$$

Esta representa el resultado medio de la autofecundación de los híbridos cuando se unen en ellos dos caracteres diferenciales. Ahora bien, en las flores individuales y en las plantas individuales, las proporciones en que aparecen las formas de las series pueden variar notablemente. Si se considera que el número de ambas clases de células huevo, en los ovarios es igual en promedio, es aleatorio cuál de las dos clases de células de polen puede fecundar a cada célula huevo individual. Por esta razón, las proporciones necesariamente están sometidas a fluctuaciones, en las que pueden darse incluso casos extremos, como se mencionó anteriormente en los experimentos sobre la forma de la semilla y el color del albumen. En estas condiciones, las proporciones numéricas poblacionales únicamente pueden estimarse mediante un promedio de la suma del mayor número posible de valores individuales. Así pues, cuanto mayor sea la cantidad de valores individuales, más se reducen los efectos aleatorios.

La serie de desarrollo para los híbridos, cuando se asocian dos caracteres diferenciales, incluye nueve formas diferentes en 16 individuos, a saber:

*AB* + *Ab* + *aB* + *ab* + 2*ABb* + 2*aBb* + 2*AaB* + 2*Aab* + 4*AaBb*

Entre los caracteres diferenciales de las plantas progenitoras originales *A*, *a* y *B*, *b*, son posibles cuatro combinaciones constantes y, por tanto, el híbrido produce las cuatro formas correspondientes de células huevo y células de polen *AB*, *Ab*, *aB*, *ab*. Cada una de estas formas participará, en promedio, cuatro veces en la fecundación, ya que se producen 16 individuos en la serie. Así, las formas participantes en la fecundación son:
  − En las células de polen:

*AB* + *AB* + *AB* + *AB* + *Ab* + *Ab* + *Ab* + *Ab* + *aB* + *aB* + *aB* + *aB* + *ab* + *ab* + *ab* + *ab*

  − En las células huevo:

$$AB + AB + AB + AB + Ab + Ab + Ab + Ab + aB + aB + aB + aB + ab + ab + ab + ab$$

En promedio, durante el proceso de la fecundación, cada forma de célula de polen se une con la misma frecuencia con cada forma de célula huevo. Por tanto, cada una de las cuatro células de polen *AB* se une una vez con cada una de las formas de células huevo *AB*, *Ab*, *aB*, *ab*. De la misma manera, se produce la unión de las otras formas de células de polen *Ab*, *aB*, *ab* con todas las demás formas de células huevo. En consecuencia, se obtiene:

$$\frac{AB}{AB} + \frac{AB}{Ab} + \frac{AB}{aB} + \frac{AB}{ab} + \frac{Ab}{AB} + \frac{Ab}{Ab} + \frac{Ab}{aB} + \frac{Ab}{ab} + \frac{aB}{AB} + \frac{aB}{Ab} + \frac{aB}{aB} + \frac{aB}{ab} + \frac{ab}{AB} + \frac{ab}{Ab} + \frac{ab}{aB} + \frac{ab}{ab}$$

O expresado de manera distinta:

$$AB + ABb + Aab + AaBb + ABb + Ab + AaBb + Aab + AaB + AaBb + aB + aBb + AaBb + Aab + aBb + ab = AB + Ab + aB + ab + 2ABb + 2aBb + 2AaB + 2Aab + 4AaBb$$

La serie de desarrollo para los híbridos, cuando se combinan tres caracteres diferenciales, puede explicarse de manera bastante similar. El híbrido produce ocho formas distintas de células huevo y células de polen: *ABC*, *ABc*, *AbC*, *Abc*, *aBC*, *aBc*, *abC*, *abc*. Cada forma de célula de polen se une nuevamente, en promedio, una vez con cada forma de célula huevo.

La ley de la combinación de los caracteres diferenciales, que rige el desarrollo de los híbridos, encuentra su fundamento y explicación en el principio de que el híbrido produce células huevo y células de polen que se corresponden, en igual número, con todas las formas constantes que surgen de la combinación de los caracteres que se unen durante la fecundación.

**10. Experimentos con Híbridos de otras Especies de Plantas**

Un objetivo futuro de investigación será determinar si la ley del desarrollo que se ha encontrado para *Pisum* también es válida para híbridos de otras plantas. Con este propósito, se han iniciado recientemente varios experimentos. Puesto que se han completado dos experimentos más pequeños con especies de *Phaseolus*, los resultados se detallan a continuación.

Un experimento con *Phaseolus vulgaris* y *Ph. nanus* L. proporcionó resultados perfectamente concordantes. *Ph. nanus* tenía, junto con el tallo enano, vainas verdes e hinchadas. Sin embargo, *Ph.*

*vulgaris* tenía un tallo de 10 a 12 pies [de 3 a 3,7 metros] de altura, y vainas de color amarillo que se contrajeron en el momento de la maduración.

Las proporciones numéricas en las que aparecieron las diferentes formas, en las distintas generaciones, fueron las mismas que en *Pisum*. El desarrollo de las combinaciones constantes también obedeció la ley de la combinación simple de caracteres, exactamente como en el caso de *Pisum*. Los resultados fueron los siguientes:

| Combinación constante | Tallo | Color de la vaina inmadura | Forma de la vaina madura |
|---|---|---|---|
| 1 | Largo | Verde | Hinchada |
| 2 | Largo | Verde | Con Constricciones |
| 3 | Largo | Amarillo | Hinchada |
| 4 | Largo | Amarillo | Con Constricciones |
| 5 | Corto | Verde | Hinchada |
| 6 | Corto | Verde | Con Constricciones |
| 7 | Corto | Amarillo | Hinchada |
| 8 | Corto | Amarillo | Con Constricciones |

El color verde de la vaina, su forma hinchada y el tallo largo fueron, como en *Pisum*, caracteres dominantes.

Otro experimento, con dos especies muy diferentes de *Phaseolus*, únicamente proporcionó resultados parciales. Se empleó *Ph. nanus* L. como planta progenitora de semillas porque es una especie con caracteres muy constantes, esto es, posee flores blancas en racimos cortos y pequeñas semillas blancas en vainas rectas, hinchadas y lisas. En cambio, *Ph. multiflorus* W. actuó como planta progenitora de polen, con un tallo alto y enrollado, flores de color rojo púrpura en racimos muy largos, vainas rugosas y dobladas en forma de hoz, y semillas grandes con motas negras sobre un fondo de color melocotón y rojo sangre.

Los híbridos tenían mayor semejanza con el progenitor productor del polen, pero las flores aparecían coloreadas con menos intensidad. Su fecundidad fue muy limitada porque, de 17 plantas –que, en conjunto, formaron muchos centenares de flores –, solo se obtuvieron 49 semillas en total. Estas semillas eran de tamaño medio y conservaban las motas negras de *Ph. multiflorus* sobre un fondo cuyo color tampoco era esencialmente diferente. Al año siguiente, se produjeron 44 plantas, de las cuales solo 31 llegaron a la fase de floración. Todos los caracteres de *Ph. nanus*, que habían quedado latentes en los híbridos, reaparecieron en varias combinaciones. No obstante, sus proporciones, en relación con las plantas dominantes, fluctuaron mucho debido al pequeño número de plantas experimentales. En ciertos caracteres, como la longitud del tallo y la forma de la vaina, la proporción fue casi exactamente de 1:3, como en *Pisum*.

Los resultados de este experimento pueden ser muy limitados para determinar las proporciones numéricas en las que aparecen las diferentes formas. Ahora bien, dichos resultados presentan el fenómeno de una apreciable transformación del color en las flores y en las semillas de los híbridos. Como ya sabemos, en *Pisum*, los caracteres del color de la flor y del color de la semilla aparecen sin cambios en la primera generación y en las subsiguientes, y la progenie de los híbridos porta exclusivamente uno u otro de los dos caracteres parentales originales. Pero tal comportamiento no se observa en este experimento. Es cierto que las flores blancas y el color de la semilla de *Ph. nanus* parecían iguales en un espécimen bastante fértil en la primera generación de los híbridos, pero las otras 30 plantas desarrollaron flores con gradaciones del color desde el rojo púrpura al violeta pálido. El color de la cubierta de las semillas no era menos variado que el de las flores. Ninguna planta pudo considerarse completamente fértil: muchas no produjeron ningún fruto y, en otras, las vainas se desarrollaron solo a partir de las últimas flores y nunca maduraron. Se obtuvieron semillas bien desarrolladas únicamente de 15 plantas. La mayor tendencia a la infertilidad se presentó en las formas con flores predominantemente rojas, en las que solo se produjeron cuatro semillas maduras de 16 plantas. En estas, tres semillas tenían marcas parecidas a las de *Ph. multiflorus*, pero con un color de fondo más o menos pálido; la cuarta planta produjo una sola semilla con un color marrón claro. Las formas con flores predominantemente de color violeta tenían semillas de color marrón oscuro, negro pardas o completamente negras.

El experimento continuó durante dos generaciones más en condiciones igualmente desfavorables, ya que, incluso entre la progenie de plantas bastante fértiles, aparecieron de nuevo algunas que eran menos fértiles o completamente estériles. No aparecieron colores de flores y semillas distintos a los mencionados. Las formas que contenían uno o más caracteres recesivos en la primera generación criada de los híbridos permanecieron constantes con esos caracteres sin excepción. Entre aquellas plantas que adquirieron flores violetas y semillas marrones o negras: (1) algunas no mostraron cambios en el color de las flores y de las semillas en las siguientes generaciones; (2) la mayoría, además de una progenie completamente similar, produjo algunas plantas con flores blancas y la cubierta de las semillas también blanca. Las plantas de flores rojas permanecieron tan poco fértiles que no puede afirmarse nada con certeza sobre su desarrollo posterior.

A pesar de las muchas dificultades que tuvieron que afrontarse para realizar estas observaciones, este experimento muestra, al menos, que el desarrollo de los híbridos sigue la misma ley que en *Pisum* en relación con aquellos caracteres correspondientes a la forma de la planta. Respecto de los caracteres del color, parece difícil encontrar una concordancia substancial. Aparte del hecho de que de la unión de un color blanco y un rojo púrpura surge toda una gama de colores, desde el púrpura hasta el violeta pálido y el blanco, es significativo que, de 31 plantas que florecieron, solo una produjo el carácter recesivo de color blanco; mientras que, en *Pisum*, esto ocurre, en promedio, en una planta de cada cuatro.

No obstante, incluso estos resultados enigmáticos podrían explicarse, probablemente, con las leyes que son válidas para *Pisum*. Para ello, se puede suponer que el color de la flor y de la semilla de *Ph. multiflorus* son un complejo de dos o más colores completamente independientes que, individualmente, se comportan como otros caracteres constantes de una planta. Si el color de la flor *A* estuviera compuesto por los caracteres independientes $A_1 + A_2 + ...$, que crearan la impresión de un color rojo púrpura, entonces, mediante la fecundación con el carácter diferencial del color blanco, *a*, se producirían las combinaciones híbridas $A_1a + A_2a + ...$, y se esperaría un comportamiento similar con el color correspondiente de las cubiertas de las semillas. De acuerdo con la suposición establecida anteriormente, cada una de estas combinaciones de colores híbridos sería independiente y, por tanto, se desarrollaría de manera completamente independiente de las demás. Así pues, al combinar las series de desarrollo individuales debe surgir una serie completa de color. Si, por ejemplo,

$$A = A_1 + A_2,$$

entonces, los híbridos $A_1a$ y $A_2a$ formarían las series de desarrollo

$$A_1 + 2A_1a + a$$
$$A_2 + 2A_2a + a$$

Los miembros de esta serie pueden presentarse en nueve combinaciones diferentes, y cada una de ellas representa otro color:

| | | | | | |
|---|---|---|---|---|---|
| 1 | $A_1 A_2$ | 2 | $A_1 a A_2$ | 1 | $A_2 a$ |
| 2 | $A_1 A_2 a$ | 4 | $A_1 a A_2 a$ | 2 | $A_2 a a$ |
| 1 | $A_1 a$ | 2 | $A_1 a a$ | 1 | $a a$ |

Los cifras asumidas para las combinaciones individuales indican también cuántas plantas con el color correspondiente pertenecen a la serie. Como esa suma es 16, todos los colores, en promedio, se distribuyen en las 16 plantas, aunque, como muestra la propia serie, en proporciones desiguales.

Si el desarrollo de los colores realmente se produjera de esta manera, podría explicarse el caso señalado anteriormente, a saber, que el color blanco de las flores y de las cubiertas de las semillas aparezca solo una vez entre 31 plantas de la primera generación. Este color solo aparece una vez en la serie. Por tanto, el color blanco podría desarrollarse, en promedio, una vez por cada 16 plantas; y, con tres caracteres de color, solo una vez por cada 64 plantas.

No debe olvidarse, sin embargo, que la explicación aquí propuesta se basa únicamente en una mera suposición que no tiene más apoyo que el resultado, muy imperfecto, del experimento que se acaba de describir. Por supuesto, sería una labor meritoria seguir el desarrollo del color en híbridos con experimentos similares, ya que es probable que, de esta manera, se llegue a comprender la extraordinaria multitud de colores en nuestras flores ornamentales.

En este punto, poco más se sabe con certeza, aparte de que el color de la flor, en la mayoría de las plantas ornamentales, es un carácter extremadamente variable. A menudo se ha expresado la opinión de que la estabilidad de la especie se ha visto alterada, en gran medida o por completo, a causa del cultivo. Hay una tendencia común a referirse al desarrollo de las formas cultivadas como si procediera sin reglas y aleatoriamente; de hecho, el color de las plantas ornamentales generalmente se cita como un ejemplo de inestabilidad. Sin embargo, no es evidente el motivo por el que la mera transferencia al suelo de un jardín debería resultar en una revolución drástica y persistente en el organismo de la planta.

Nadie podría afirmar seriamente que el desarrollo de las plantas en campo abierto se rige por leyes diferentes a las que gobiernan su desarrollo en el suelo de un jardín. Aquí, como allá, deben aparecer variaciones tipológicas si se cambian las condiciones de vida de una especie, y esta tiene la capacidad de adaptarse a las nuevas condiciones. Ahora bien, se puede admitir abiertamente que, mediante el cultivo, se favorece la aparición de nuevas variedades y que, mediante la intervención humana, se conservan muchas variaciones que habrían fracasado en condiciones naturales.

Consecuentemente, nada justifica la suposición de que la disposición de una especie a desarrollar un gran número de nuevas variedades cause que dicha especie pierda pronto toda estabilidad y que su progenie se fragmente en una serie infinita de formas altamente variables. Si el cambio en las condiciones de la vegetación fuera la única causa de la variabilidad, estaría justificado esperar que aquellas plantas domesticadas, cultivadas casi en las mismas condiciones durante siglos, hubieran adquirido estabilidad. Como se sabe, este no es el caso, ya que, especialmente en estas plantas, se encuentran no solo las formas más variadas sino también las más variables. Únicamente las leguminosas, como *Pisum*, *Phaseolus* y *Lens*, cuyos órganos de fecundación están protegidos por la quilla, constituyen una excepción destacable. Incluso para estas leguminosas, han surgido numerosas variedades durante más de 1000 años de cultivo en las más diversas condiciones. Sin embargo, bajo las mismas condiciones permanentes de vida, conservan una estabilidad similar a la de las especies silvestres.

Es más que probable que, en lo que se refiere a la variabilidad de las plantas cultivadas, exista un factor que, hasta ahora, ha recibido poca atención. Varios experimentos nos fuerzan a concluir que nuestras plantas ornamentales, con pocas excepciones, son miembros de diferentes series de híbridos, cuyo desarrollo posterior, regido por una ley, se modifica y se interrumpe a causa de frecuentes uniones entre ellas.

Debe enfatizarse que las plantas cultivadas normalmente se crían en gran número y una al lado de la otra, circunstancias que favorecen la fecundación entre las variedades existentes y entre las propias especies. Esta suposición encuentra apoyo en el hecho de que, entre la gran multitud de formas variables, siempre se hallan individuos que permanecen constantes en uno u otro carácter si se previene cuidadosamente toda influencia externa. Estas formas se desarrollan exactamente igual que ciertos miembros de las series complejas de híbridos.

Incluso con el más sensible de todos los caracteres, el del color, al observador atento no se le escapa que, en las formas individuales, su tendencia a la variabilidad se presenta en grados muy diversos. Entre las plantas que descienden de una fecundación espontánea: (1) a menudo hay algunas cuya progenie varía ampliamente en la constitución y disposición de los colores; (2) en otras plantas, sin embargo, su progenie muestra pocas diferencias; y (3) cuando se examina un mayor número de individuos, se encuentran algunas plantas que transmiten sin cambios el color de las flores a su progenie.

Como modelo ilustrativo de lo que se ha descrito en el párrafo anterior, sirven las especies cultivadas de *Dianthus*. Un espécimen de flores blancas de *Dianthus caryophyllus*, que a su vez procedía de una variedad de flores blancas, se aisló en un invernadero durante su período de floración. Las numerosas semillas que se obtuvieron de él produjeron plantas con exactamente el mismo color blanco en las flores. Un resultado similar se produjo con una variedad de flores rojas algo teñidas de violeta, y con una variedad de flores blancas y rojas a rayas. En cambio, otras muchas plantas, que fueron protegidas de la misma manera, produjeron una progenie con colores y marcas más o menos variados.

Quienquiera que examine los colores en las plantas ornamentales que surgen de fecundaciones similares llegará fácilmente a la convicción de que, en este caso, el desarrollo también se rige por una ley concreta que, posiblemente, puede expresarse como la *combinación de varios caracteres de color independientes*.

**11. Discusión**

Aunque pueda carecer de interés, a continuación se comparan las observaciones extraídas del estudio sobre *Pisum* con los resultados de las investigaciones que realizaron Kölreuter y Gärtner. Según la opinión de ambas autoridades en la materia, la apariencia externa de los híbridos, o bien adquiere una forma intermedia entre aquellas de los parientes originales, o bien se asemeja más al tipo de uno de los progenitores; en este último caso, el parecido es a veces tan estrecho que apenas pueden diferenciarse. Generalmente, si la fecundación se efectúa por autopolinización, las semillas producidas son de formas diferentes, distintas del tipo normal. Como regla general, puede afirmarse que tras una fecundación: (1) la mayoría de los individuos adquiere una forma híbrida; (2) otros se asemejan a la planta progenitora que produce las semillas; y (3) algún que otro individuo se parece más a la planta progenitora que produce el polen. Sin embargo, esto no ocurre así en todos los híbridos sin excepción. De hecho, en la progenie de algunos híbridos, puede darse una de las siguientes situaciones: o bien (1) una parte de su progenie se asemeja a una de las plantas parentales, y la otra parte se parece más a la otra planta parental; o bien (2) toda la progenie tiende a asemejarse a una de las plantas parentales; o bien (3) la progenie sigue manteniendo perfectamente su similitud con el híbrido, y esta similitud continúa constante en la descendencia posterior. Los híbridos de variedades se comportan como los híbridos de especies, pero los híbridos de variedades poseen una variabilidad en sus atributos aún mayor y una tendencia más pronunciada a volver a las formas parentales originales.

En relación con los atributos de los híbridos y su desarrollo, la concordancia con las observaciones realizadas en *Pisum* es inequívoca, pero no es así con los casos excepcionales que hemos mencionado. El mismo Gärtner admite que la determinación precisa de si una forma es más similar a uno u otro de los dos progenitores es, a menudo, extremadamente difícil porque depende, en gran medida, de la visión subjetiva del observador. No obstante, otra circunstancia podría haber contribuido a resultados fluctuantes e inciertos a pesar de la observación y la comparación más cuidadosas. Para los experimentos se utilizaron plantas que, en su mayoría, se consideran buenas especies y están diferenciadas por un gran número de caracteres. Ahora bien, cuando se trata de caracteres bien definidos, la cuestión se reduce a afirmar si la semejanza es mayor o menor. Sin embargo, también se deben tener en cuenta aquellos caracteres que, a pesar de que resultan difíciles de describir con palabras, confieren a las formas un aspecto peculiar, como bien sabe todo especialista en plantas.

Así pues, si suponemos que el desarrollo de los híbridos se rige por las leyes aplicables a *Pisum*, entonces, la serie de cada experimento individual debe incluir muchas formas porque, como sabemos, el número de términos aumenta en potencias de 3 con respecto al número de caracteres diferenciales. Por tanto, con un número relativamente pequeño de plantas experimentales, el resultado podría ser solo aproximadamente correcto y, en casos individuales, podría desviarse significativamente. Por ejemplo, si dos progenitores originales difieren en siete caracteres, y se cultivan de 100 a 200 plantas a partir de las semillas de sus híbridos para evaluar el grado de parentesco entre la progenie, es obvio lo incierta que puede llegar a ser la conclusión que se extraiga de los resultados. Es decir, para siete caracteres diferenciales, la serie combinatoria constaría de 2187 formas diferentes, que incluirían a 16384 individuos; por tanto, en los resultados, podría predominar una relación u otra, dependiendo de si el azar, en la mayoría de los casos, ha presentado al observador una forma o la otra.

Por otra parte, sabemos que los caracteres dominantes se transmiten al híbrido sin alteraciones o casi sin cambios. Entonces, si estos caracteres dominantes aparecen entre los caracteres diferenciales, en los términos de las series de desarrollo debe siempre predominar el progenitor que posee mayor número de caracteres dominantes. A este respecto, en el experimento sobre *Pisum* con tres caracteres diferenciales, al que se aludió anteriormente, todos los caracteres dominantes pertenecían a la planta portadora de las semillas. Si bien los términos de la serie, conforme a su naturaleza interna, tienden por igual hacia ambas plantas parentales originales, en este experimento, el tipo del progenitor productor de las semillas fue tan preponderante que, de cada 64 plantas de la primera generación, 54 se parecían exactamente a él o diferían en un solo carácter. Dadas las circunstancias, es evidente lo arriesgado que puede ser a veces extraer conclusiones sobre el parentesco interno de los híbridos a partir de su apariencia externa.

Gärtner menciona que, en aquellos casos en los que el desarrollo fue regular, los progenitores originales no reaparecían en la descendencia de los híbridos, sino que algunos individuos se aproximaban a ellos. De hecho, con series de desarrollo muy extensas, no podría ocurrir de otra manera. Por ejemplo, con siete caracteres diferenciales, entre los más de 16000 descendientes de los híbridos, las dos formas parentales originales aparecerían solo una vez cada una. En consecuencia, no es muy probable que los dos progenitores aparezcan entre un pequeño número de plantas experimentales; no obstante, con cierta probabilidad, se puede esperar la aparición en la serie de algunas formas individuales similares a uno de ellos.

Encontramos una diferencia esencial en aquellos híbridos que permanecen constantes en su progenie y se propagan de la misma manera que las especies puras. Según Gärtner, entre estos se encuentran los siguientes híbridos, que se caracterizan por su notable fertilidad: *Aquilegia atropurpurea-canadensis*, *Lavatera pseudolbia-thuringiaca*, *Geum urbano-rivale*, algunos híbridos de *Dianthus* y, según Wichura, híbridos de especies de sauce. Esta circunstancia es especialmente importante para la historia evolutiva de las plantas porque los híbridos constantes adquieren el estatus de nuevas especies. Los más destacados observadores garantizan la veracidad de este hecho, y no puede ponerse en cuestión. Gärtner tuvo la oportunidad de seguir el desarrollo de la planta *Dianthus armeria-deltoides* hasta la décima generación, ya que se reproducía por sí misma y con regularidad en el jardín.

Con *Pisum*, los experimentos demostraron que los híbridos forman diferentes células huevo y células de polen, y que esta es la razón de la variabilidad en su progenie. Podemos suponer que sea esta también la causa de la variabilidad en el caso de otros híbridos cuya progenie se comporte de manera similar. Sin embargo, para aquellos que permanecen constantes, parece admisible la suposición de que sus células reproductoras son todas iguales y coinciden con la célula fundadora del híbrido.

De acuerdo con la opinión de renombrados fisiólogos, la reproducción en las fanerógamas se realiza mediante la unión de una célula huevo y una célula de polen que da lugar a una sola célula[2], capaz de convertirse en un organismo independiente a través de la absorción de material y la formación de nuevas células. Este desarrollo se rige por una ley constante, que se fundamenta tanto en la composición material de los elementos como en su disposición para lograr una unión viable en la célula. Si las células reproductoras son del mismo tipo y coinciden con la célula fundadora de la planta madre, entonces el desarrollo del nuevo individuo se regirá por la misma ley que se aplica a la planta madre. En cambio, si se produce una unión viable entre una célula huevo y una célula de polen de un tipo diferente, debemos suponer que, entre los elementos de ambas células que determinan sus diferencias, existe algún mecanismo de reajuste. La célula intermedia que surge se convierte en el fundamento del organismo híbrido, cuyo desarrollo sigue, necesariamente, una ley distinta a aquella que rige en las dos plantas parentales originales. Si consideramos que el reajuste es completo, en el sentido de que el embrión híbrido se forma a partir de células del mismo tipo en las que las diferencias están totalmente conciliadas y de manera permanente, entonces se puede concluir que el híbrido, como cualquier otra especie de planta autónoma, permanecerá constante en su progenie. Las células reproductoras que se formen en los ovarios y en las anteras serán del mismo tipo y coincidirán con la célula intermedia subyacente.

---

[2] En *Pisum* se demuestra, sin lugar a dudas, que debe existir una unión completa de los elementos de ambas células reproductoras para la formación del nuevo embrión. Si no fuera así, ¿cómo podría explicarse que, entre la progenie de los híbridos, ambas formas originales reaparezcan en igual número y con todas sus peculiaridades? Si la influencia de la célula huevo sobre la célula de polen fuera solo externa, esto es, si actuara únicamente a modo de nodriza, entonces el resultado de cada fecundación artificial sería que el híbrido desarrollado se parecería exclusivamente a la planta de polen, de manera exacta o muy similar. Sin embargo, los experimentos realizados hasta ahora no han probado de ninguna manera este hecho. Una evidencia fundamental de la unión completa del contenido de ambas células reside en la experiencia universalmente confirmada de que la forma del híbrido no depende de cuál de las formas originales era la planta productora de las semillas o la planta productora de polen.

En relación con aquellos híbridos cuya progenie es variable, quizás podría suponerse que se logra una conciliación entre los elementos diferenciales de la célula huevo y la célula de polen. Dicha conciliación consiste en hacer posible la formación de una célula como fundamento del híbrido, pero el reajuste de los elementos opuestos es solo temporal y no se extiende más allá de la vida de la planta híbrida. Como no se perciben cambios en la apariencia general de la planta a lo largo del período vegetativo, debemos inferir que los elementos diferenciales consiguen liberarse de su asociación forzosa solo durante el desarrollo de las células reproductoras. En la formación de estas células participan todos los elementos existentes de una manera completamente libre y equilibrada, pero los elementos diferenciales se segregan entre sí. Así, se posibilitaría la producción de tantas células huevo y células de polen como combinaciones de elementos formativos existan.

Esta diferencia esencial en el desarrollo de los híbridos se está explicando aquí como una asociación permanente o temporal de los elementos celulares diferenciales. Por supuesto, dada la escasez de datos precisos, esta explicación, que deja un amplio margen para la interpretación, puede tratarse como una hipótesis. Ahora bien, la opinión expresada encuentra justificación en las evidencias que aportan los experimentos con *Pisum*. Por una parte, el comportamiento de cada par de caracteres diferenciales, en la unión híbrida, es independiente de las otras diferencias entre las dos plantas originales. Por otra parte, el híbrido produce tantos tipos de células huevo y células de polen como combinaciones posibles hay de formas constantes. En última instancia, los caracteres distintivos de dos plantas solo pueden proceder de las diferencias en la composición y agrupación de los elementos existentes en sus células fundadoras, donde dichos elementos sostienen una interacción vital.

La validez del conjunto de leyes establecidas para *Pisum* requiere una confirmación adicional. Por tanto, sería deseable la repetición, al menos, de los experimentos más importantes, por ejemplo, el relativo a la composición de las células reproductoras de los híbridos. Puesto que en los experimentos ha existido un único observador, fácilmente puede pasar inadvertido un carácter diferencial; aunque esta diferencia parezca insignificante al principio, puede adquirir tal relevancia que no pueda obviarse en el resultado final. Igualmente, debe determinarse mediante experimentos si los híbridos variables de otras especies de plantas muestran un comportamiento completamente idéntico; no obstante, puede suponerse que, en aspectos relevantes, no puede aparecer una diferencia en lo consustancial ya que la unidad del plan evolutivo de la vida orgánica es incuestionable.

En conclusión, merecen una mención particular los experimentos realizados por Kölreuter, Gärtner y otros sobre la transformación de una especie en otra mediante fecundación artificial. Se ha dado una importancia particular a estos experimentos, y Gärtner los considera entre los "más difíciles en la producción de híbridos".

Para que una especie *A* se convirtiera en otra *B*, ambas se unían por fecundación y los híbridos producidos se fecundaban nuevamente con el polen de *B*. Luego, de sus diferentes descendientes, se seleccionaba la forma más próxima a la especie *B* y se fecundaba repetidamente con ella, y así sucesivamente hasta que, finalmente, se lograba una forma que se parecía mucho a *B* y permanecía constante en su progenie. Se procedía de esta manera, por tanto, para que la especie *A* se transformara en la otra especie *B*. El propio Gärtner realizó 30 experimentos de este tipo con plantas de los géneros *Aquilegia*, *Dianthus*, *Geum*, *Lavatera*, *Lychnis*, *Malva*, *Nicotiana* y *Oenothera*. La duración de la transformación no fue la misma para todas las especies. De hecho, tres fecundaciones fueron suficientes para algunas especies, pero otras necesitaron de cinco a seis fecundaciones; asimismo, para estas mismas especies se observaron fluctuaciones en diferentes experimentos. Gärtner atribuye esta diferencia a la circunstancia de que "el vigor tipológico con el que una especie actúa en la reproducción para cambiar y modificar el tipo materno es muy diferente en las distintas plantas, consecuentemente, los períodos de tiempo y el número de generaciones necesarios para que una especie se transforme en otra también deben variar; por tanto, la transformación de una especie en otra requiere de más o menos generaciones dependiendo de cada especie concreta". Además, el mismo observador señaló que "en los procesos de transformación también influyen el tipo y el individuo que se seleccionan para las transformaciones posteriores".

Si suponemos que el desarrollo de las formas en estos experimentos tenía lugar de manera similar a la de *Pisum*, entonces, todo el proceso de transformación podría explicarse de una manera sencilla. El híbrido forma tantas clases de células huevo como sean admisibles dadas las combinaciones constantes de sus caracteres agregados, y una de estas es siempre de la misma clase que la de las células de polen fecundantes. Por tanto, en todos estos experimentos es posible que, ya en la segunda fecundación, se obtenga una forma constante que se asemeje a la planta productora del polen. Sin embargo, el hecho de que esto realmente se produzca depende, en cada caso particular, del número de plantas experimentales y del número de caracteres diferenciales que se unen a través de la fecundación. Supongamos, por ejemplo, que las plantas seleccionadas para el experimento fueran diferentes en tres caracteres y que la especie *ABC* tuviera que transformarse en otra especie *abc* mediante la fecundación repetida con el polen de esta especie. El híbrido que surge de la primera fecundación forma ocho tipos diferentes de células huevo, a saber:

*ABC*, *ABc*, *AbC*, *aBC*, *Abc*, *aBc*, *abC*, *abc*.

Estos se combinan nuevamente con las células de polen *abc* en el segundo año experimental, y se obtiene la serie:

*AaBbCc* + *AaBbc* + *AabCc* + *aBbCc* + *Aabc* + *aBbc* + *abCc* + *abc*.

Debido a que la forma *abc* aparece una vez en la serie de ocho miembros, es menos probable que esté ausente entre las plantas experimentales, incluso si se criaran en menor número, y la transformación se completaría después de solo dos fecundaciones. Si, por casualidad, no se produjera, sería necesario repetir la fecundación con una de las combinaciones relacionadas más cercanas: *Aabc*, *aBbc*, *abCc*. Se hace evidente que, cuanto menor sea el número de plantas experimentales y mayor el número de caracteres diferenciales en los dos progenitores originales, más tiempo tendrá que prolongarse dicho experimento; además, en la misma especie puede fácilmente ocurrir un retraso de una o incluso de dos generaciones, como observó Gärtner. La transformación de especies muy divergentes bien puede terminarse en el quinto o sexto año experimental porque el número de células huevo diferentes que se forman en el híbrido aumenta, en potencias de 2, con el número de caracteres diferenciales.

Gärtner encontró, a través de la repetición de los experimentos, que el tiempo de transformación para algunas especies es diferente, por lo que, a menudo, la especie *A* se puede convertir en otra *B* una generación antes que la especie *B* en *A*. También deduce de esta evidencia que la opinión de Kölreuter no está completamente fundamentada, según la cual "las dos naturalezas en el híbrido están en perfecto equilibrio". Sin embargo, parece que Kölreuter no merece tal crítica porque, más bien, Gärtner pasó por alto un factor importante sobre el que él mismo llama la atención en otro lugar, esto es, "influye qué individuo se selecciona para las transformaciones posteriores". Los experimentos a este respecto realizados con dos especies de *Pisum* indican que, en lo referente a la elección de los individuos más apropiados para ulteriores fecundaciones, puede haber una gran diferencia según cuál de las dos especies se transforma en la otra. Las dos plantas experimentales diferían en cinco caracteres, la especie *A* poseía todos los caracteres dominantes, y la otra especie *B*, todos los caracteres recesivos. Para la transformación entre especies, *A* se fecundó con el polen de *B*, y, a la inversa, *B* se fecundó con el polen de *A*. Luego, se repitió el mismo proceso al año siguiente con ambos híbridos. Con el primer experimento, ***B/A***, hubo en el tercer año experimental 87 plantas con todas las 32 formas posibles, que estaban disponibles para la selección de individuos para la fecundación posterior. Con el segundo experimento, ***A/B***, resultaron 73 plantas, que, en su apariencia general, eran completamente idénticas a la planta progenitora productora del polen, pero, según su naturaleza interna, eran necesariamente tan diferentes como las formas del otro experimento. Consecuentemente, solo fue posible una selección definida en el primer experimento; en el segundo experimento, hubo que descartar algunas plantas al azar. De estas últimas, solo una parte de las flores se fecundaron con el polen de *A*, y las demás se dejaron autofecundarse. Como mostró el cultivo del año siguiente, de cada cinco plantas utilizadas para la fecundación en los dos experimentos, concordaron con el progenitor productor del polen las siguientes proporciones de plantas:

| Primer experimento | Segundo experimento | |
|---|---|---|
| 2 plantas | — | en todos los caracteres |
| 3 plantas | — | en 4 caracteres |
| — | 2 plantas | en 3 caracteres |
| — | 2 plantas | en 2 caracteres |
| — | 1 planta | en 1 carácter |

En el primer experimento, por tanto, se completó la transformación; en el segundo, que no se continuó, habrían sido necesarias, probablemente, dos fecundaciones adicionales.

Aunque no sea frecuente el caso en que los caracteres dominantes pertenezcan exclusivamente a una u otra planta parental original, sí influye cuál de las dos plantas posee el mayor número de caracteres dominantes. Si la mayoría proviene de la planta productora de polen, la selección de formas para la fecundación posterior proporcionará un menor grado de certeza en comparación con el caso contrario. Además, se requerirá más tiempo para la transformación si el experimento se considera terminado cuando se obtenga una forma que no solo se parezca a la planta productora de polen en sus atributos, sino que también permanezca constante en su progenie.

Los resultados que Gärtner obtuvo con sus experimentos de transformación explican que este se haya opuesto a la opinión de aquellos naturalistas que cuestionan la estabilidad de las especies de plantas y asumen una evolución continua en sus variedades. Él ve en la transformación completa de una especie en otra la evidencia inequívoca de que una especie tiene límites fijos, más allá de los cuales no puede cambiar. Aunque no se pueda otorgar una validez incondicional a esta opinión, en los experimentos realizados por Gärtner se encuentra, sin embargo, una confirmación, con respecto a la suposición expresada anteriormente sobre la variabilidad de las plantas cultivadas, que merece resaltarse.

Entre las especies experimentales se hallaban plantas cultivadas, como *Aquilegia atropurpurea* y *canadensis*; *Dianthus caryophyllus*, *chinensis* y *japonicus*; y *Nicotiana rustica* y *paniculata*. Estas especies tampoco habían perdido nada de su autonomía después de cuatro o cinco uniones híbridas.